\documentclass[pra,twocolumn,superscriptaddress,showpacs,amsmath,amstex,amssymb,citeautoscript]{revtex4-1}

\usepackage[T1]{fontenc}
\usepackage[utf8]{inputenc}
\usepackage{lipsum}
\usepackage{amsmath}
\usepackage{amssymb}
\usepackage{bbm}
\usepackage[multiple]{footmisc}
\usepackage{braket}
\usepackage{xcolor}
\usepackage{pifont}
\newcommand{\xmark}{\ding{55}}%
\usepackage[mathscr]{euscript}
\usepackage[shortlabels]{enumitem}
\usepackage[justification=justified]{subcaption}
\usepackage{graphicx}
\usepackage{lipsum}
\allowdisplaybreaks
\usepackage{float}
\usepackage{graphicx}
\usepackage{dsfont}
\usepackage{comment}
\usepackage[colorlinks=true]{hyperref}  
\hypersetup{
    bookmarks=true,         
    unicode=false,          
    pdftoolbar=true,        
    pdfmenubar=true,        
    pdffitwindow=false,     
    pdfstartview={FitH},    
    pdftitle={Altermagnetic superconducting diode effect},    
    pdfauthor={Banerjee and Scheurer},     
    pdfsubject={},   
    pdfcreator={},   
    pdfproducer={}, 
    pdfkeywords={} {} {}, 
    pdfnewwindow=true,      
    colorlinks=true,       
    linkcolor=blue, 
    citecolor=blue,        
    filecolor=magenta,      
    urlcolor=blue           
}

\newcommand{\equref}[1]{Eq.~(\ref{#1})}
\newcommand{\equsref}[2]{Eqs.~(\ref{#1}) and (\ref{#2})}
\newcommand{\secref}[1]{Sec.~\ref{#1}}
\newcommand{\figref}[1]{Fig.~\ref{#1}}
\newcommand{\refcite}[1]{Ref.~\onlinecite{#1}}

\newcommand{\tableref}[1]{Table~\ref{#1}}
\newcommand{\appref}[1]{Appendix~\ref{#1}}

\newcommand{\pdagger}{{\phantom{\dagger}}}

\renewcommand{\vec}[1]{\boldsymbol{#1}}

\definecolor{wrongultramarine}{rgb}{1,0.5,0}

\begin{document}

\title{Altermagnetic superconducting diode effect}

\author{Sayan Banerjee}
\affiliation{Institute for Theoretical Physics III, University of Stuttgart, 70550 Stuttgart, Germany}

\author{Mathias S.~Scheurer}
\affiliation{Institute for Theoretical Physics III, University of Stuttgart, 70550 Stuttgart, Germany}

\begin{abstract}
Non-reciprocal superconductivity, also known as the superconducting diode effect, has been extensively studied in the presence of a magnetic field or some form of ferromagnetic order breaking time-reversal symmetry. We here show that another class of magnetic order known as altermagnetism, which also breaks time-reversal symmetry but does not exhibit a finite net magnetic moment, can also give rise to a superconducting diode effect. Whether this is the case depends on the combination of the system's point group and altermagnetic order parameter which we explore systematically for two-dimensional crystalline systems. If the superconducting electrons are in a centrosymmetric crystalline environment, an electric field $E_z$ (or other sources of inversion symmetry breaking) can be used to turn on and tune the non-reciprocity, yielding an electric-field tunable diode effect; there are also non-centrosymmetric point groups, which are not reached by applying $E_z \neq 0$ in a centrosymmetric crystal, but  still allow for an altermagnetic order parameter with non-reciprocal superconductivity. Depending on the residual magnetic point group, the zeros of the critical current asymmetry, $J_c(\hat{\vec{n}}) - J_c(-\hat{\vec{n}})$, are pinned along high-symmetry crystalline directions $\hat{\vec{n}}=\hat{\vec{e}}_j$ or are free to rotate in the plane of the system. In some cases, the zeros can be rotated by tuning the electric field $E_z$.  We discuss all of these phenomena both on the general level using exact symmetry arguments and more explicitly by constructing and solving minimal lattice models. 
We provide experimental setups to realize the altermagnetic superconducting diode effect.

\end{abstract}

\maketitle

\section{Introduction}

The broad class of magnetically ordered states referred to as ``altermagnets’’ (AMs) has recently become a subject of notable interest and has motivated a significant amount of theoretical \cite{yuan_giant_2020,smejkal_beyond_2022,smejkal_emerging_2022,ahn_antiferromagnetism_2019,bhowal_magnetic_2022,cuono_orbital-selective_2023,guo_spin-split_2023,maier_weak-coupling_2023,mazin_altermagnetism_2023,oganesyan_quantum_2001,sato_altermagnetic_2023,steward_dynamic_2023,turek_altermagnetism_2022,ouassou_dc_2023,mazin_prediction_21,hayami_momentum_19,smejkal_crystal_20,brekke_two-dimensional_2023,das_realizing_2023,fakhredine_interplay_2023,fang_quantum_2023,leeb_spontaneous_2023,mazin_induced_2023,smejkal_chiral_2023,sun_spin_2023,gao_ai-accelerated_2023,beenakker_phase-shifted_2023,thermal_transport,RafaelsPaper,andreev,zhang2023finitemomentum,brekke_two-dimensional_2023,giil2023superconductoraltermagnet,majorana1,majorana2,majorana3,orientation_altermagnet,antonenko2024mirror,yu2024altermagnetism,wei2023gapless,mcclarty2023landau} and experimental \cite{aoyama_piezomagnetic_2023,bai_efficient_2023,krempasky_altermagnetic_2023,lee_broken_2023,reimers_direct_2023,feng_anomalous_2022,bose_tilted_2022} studies, probing into its nature and emergence. Just as with any form of magnetic order, time-reversal symmetry is broken in an AM. However, as opposed to ferromagnets, AMs do not exhibit a finite net magnetic moment, which is guaranteed by a magnetic point symmetry. This means that the state is invariant under the product $\Theta g_m$ of time-reversal $\Theta$ and a unitary lattice symmetry $g_m$. Unlike a collinear antiferromagnet, $g_m$ is not lattice translation for an AM, where the reversal of all local magnet moments resulting from application of $\Theta$ cannot be undone by a lattice translation $T_{\vec{R}}$. Instead, $g_m$ is a point group symmetry such as a rotation or reflection symmetry of the underlying crystalline lattice.

While a Pomeranchuk instability \cite{pomeranchuck,wu_fermi_2007} in the $l=2$ (or higher $l$) spin-triplet channel of a Fermi liquid 
falls into the category of AMs, a lot of interesting novel consequences have recently been uncovered. 
For instance, since $g_m$ is not translation, AMs can exhibit an anomalous Hall effect, even without net magnetic moment, unlike antiferromagnets where $\Theta T_{\vec{R}}$ does not allow for a finite Hall response \cite{feng_anomalous_2022,sato_altermagnetic_2023}. Furthermore, the spin texture of the Bloch states in the presence of an AM can lead to a Zeeman splitting with protected nodal lines \cite{RafaelsPaper}. 
Further consequences have been explored in thermal transport \cite{thermal_transport}, in Josephson junctions \cite{beenakker_phase-shifted_2023,ouassou_dc_2023}, and other tunnel junction setups \cite{chi2023crystal}. What is more, the effect of AMs on other coexisting interaction-induced orders is very rich; in particular, superconductivity \cite{chakraborty_zero-field_2023,zhang2023finitemomentum,brekke_two-dimensional_2023,giil2023superconductoraltermagnet,majorana1,majorana2,majorana3,orientation_altermagnet,wei2023gapless} has been explored, sparking off broader questions regarding the nature of strongly correlated materials exhibiting superconductivity and magnetism beyond the conventional paradigm of ferromagnetism and antiferromagnetism.

Concurrently, non-reciprocal transport in superconductors and superconducting junctions, particularly the superconducting diode effect (SDE), has garnered various experimental \cite{diode0,ando_observation_2020,lyu_superconducting_2021,du_superconducting_2023,sundaresh_diamagnetic_2023,kealhofer_anomalous_2023,hou_ubiquitous_2023,chen_superconducting_2023,gupta_superconducting_2022,AnotherJosephson,banerjee_phase_2023,pal_josephson_2022,kim_intrinsic_2023,gupta_superconducting_2022,MoreJosephson,ParadisoNbSe2,wu_field-free_2022,diez-merida_magnetic_2021,golod_demonstration_2022,chiles_non-reciprocal_2022-1,zhang_reconfigurable_2023,shin_magnetic_2021,jeon_zero-field_2022,kealhofer_anomalous_2023,hou_ubiquitous_2023,lin_zero-field_2022,anwar_spontaneous_2022,narita_field-free_2022,gutfreund_direct_2023,cuprate_diode} and theoretical \cite{daido_superconducting_2022,daido_intrinsic_2022,yuan_supercurrent_2022,he_phenomenological_2022,ilic_theory_2022,scammell_theory_2022,zinkl_symmetry_2022,PhysRevX.12.041013,he_supercurrent_2022,PhysRevB.106.L140505,jiang_field-free_2022,kokkeler_field-free_2022,debnath2024gatetunable,chazono_piezoelectric_2022,vodolazov_superconducting_2005,de_picoli_superconducting_2023,kochan_phenomenological_2023,ikeda_intrinsic_2022,tanaka_theory_2022,wang_symmetry_2022,haenel_superconducting_2022,legg_parity_2023,cuozzo_microwave-tunable_2023,souto_josephson_2022,cheng_josephson_2023,steiner_diode_2022,costa_microscopic_2023,wei_supercurrent_2022,legg_superconducting_2022,karabassov_hybrid_2022,josepheson_valleypol,PhysRevLett.130.126001,nunchot2024chiral,daido2023unidirectional,souto2023tuning,cayao} attention. 
The SDE refers to a difference in the critical current $J_c(\hat{\vec{n}})$ in opposite directions, $\hat{\vec{n}}$ and $-\hat{\vec{n}}$. In that case, applying a current $J$ with $\min  (J_c(\pm\hat{\vec{n}})) < J < \max  (J_c(\pm\hat{\vec{n}}))$, leads to superconducting transport along one and resistive behavior along the opposite direction, yielding a superconducting analogue of a semi-conducting diode. On top of fundamental theoretical questions, the rich space of potential applicability of such non-reciprocal responses on modern-day electronics and devices renders the SDE an exciting topic to delve into from an experimental and technological standpoint.  

Importantly, time-reversal symmetry $\Theta$ implies $J_{c}(\hat{\vec{n}}) = J_{c}(-\hat{\vec{n}})$ which therefore needs to be broken to obtain a finite SDE. So far, external magnetic fields \cite{ando_observation_2020,du_superconducting_2023,hou_ubiquitous_2023,chen_superconducting_2023,gupta_superconducting_2022,AnotherJosephson,banerjee_phase_2023,pal_josephson_2022,kim_intrinsic_2023,MoreJosephson,ParadisoNbSe2,lyu_superconducting_2021,sundaresh_diamagnetic_2023,kealhofer_anomalous_2023,diez-merida_magnetic_2021,daido_superconducting_2022,daido_intrinsic_2022,yuan_supercurrent_2022,he_phenomenological_2022,ilic_theory_2022,he_supercurrent_2022,de_picoli_superconducting_2023,kochan_phenomenological_2023,ikeda_intrinsic_2022,cheng_josephson_2023,legg_superconducting_2022,PhysRevLett.130.126001,nunchot2024chiral}, proximity-induced magnetism \cite{shin_magnetic_2021,jeon_zero-field_2022,narita_field-free_2022}, and interaction-induced order parameters in the same electron liquid \cite{lin_zero-field_2022,scammell_theory_2022,banerjee_enhanced_2024,josepheson_valleypol} have been discussed as possible causes for the broken time-reversal symmetry inducing the SDE. All of these  $\Theta$-breaking orders fall under the category of (spin and/or orbital) ferromagnetism and are associated with a non-zero net magnetic moment. Due to its residual magnetic symmetry $\Theta T_{\vec{e}_{x,y}}$, with nearest-neighbor vectors $\vec{e}_{x,y}$, collinear antiferromagnetism as the only cause of symmetry reduction cannot stabilize the SDE. This therefore begs the question if and under which conditions AMs can induce a SDE in an otherwise symmetry-unbroken superconductor and what the consequences are of the non-trivial magnetic point symmetries of the AM on the critical current $J_c(\hat{\vec{n}})$. This is the scope of our present work.

To this end, we here systematically analyze, using a combination of symmetry arguments and explicit model calculations, which of the possible AMs in two-dimensional (2D) crystalline systems can yield a critical current asymmetry $J_c(\hat{\vec{n}}) - J_c(-\hat{\vec{n}}) \neq 0$. Indeed, there are several candidate AMs where a SDE is possible despite the lack of net magnetization. We will show that, depending on the residual magnetic point group, the zeros of the current asymmetry $J_c(\hat{\vec{n}}) - J_c(-\hat{\vec{n}})$ are either aligned with high-symmetry crystalline directions, can be rotated by an electric field, or are not at all pinned to the crystalline axes.  

The remainder of the manuscript is organized as follows. In \secref{general}, we describe the method of calculating $J_{c}(\hat{\vec{n}})$ and discuss the general constraints on its directional dependence and on its asymmetry resulting from magnetic point symmetries. We further systematically go through all AM order parameters in 2D single-band models, study whether they induce an SDE and, if so, how the current asymmetry $J_c(\hat{\vec{n}}) - J_c(-\hat{\vec{n}})$ depends on $\hat{\vec{n}}$.  
In \secref{ModelStudies}, we demonstrate the AM-induced SDE on a case-by-case basis by constructing and solving explicit lattice models, detailing the distinctive characters in each of the cases. Finally, \secref{Conclusion} summarizes our findings.

\section{General considerations}
\label{general}
Before presenting explicit calculations, we start by discussing general constraints on the AM-induced SDE, resulting from symmetries. We will see under which conditions the SDE is forced to vanish along high-symmetry directions and systematically go through all possible AM order parameters of 2D systems. 

\subsection{Diode effect efficiency}\label{diode_efficiency}
To set the stage for our symmetry discussion, we begin by first outlining the basic procedure for the calculation of the critical current. Denoting the superconducting order parameter for Cooper pairs with center-of-mass momentum $\vec{q}$ by $\Delta_{\vec{q}}$, we expand the change of the free-energy as a result of superconductivity $\delta \mathcal{F}_{\text{S}} := \mathcal{F}[\Delta_{\vec{q}}] - \mathcal{F}[0]$ up to quartic order in $\Delta_{\vec{q}}$,
\begin{equation}
    \delta \mathcal{F}_{\text{S}} \sim \sum_{\vec{q}} a^{\text{S}}_{\vec{q}} \, |\Delta_{\vec{q}}|^2 + b^{\text{S}} \sum_{\vec{q}_i} \Delta^*_{\vec{q}_1}\Delta^*_{\vec{q}_2}\Delta^{\phantom{*}}_{\vec{q}_3}\Delta^{\phantom{*}}_{\vec{q}_4} \delta_{\vec{q}_1+\vec{q}_2,\vec{q}_3+\vec{q}_4},
    \label{eq:gl_free}
\end{equation}
where we neglected the momentum dependence of the quartic term. 
The equilibrium superconducting state is found by minimizing $\delta \mathcal{F}_{\text{S}}$. Restricting the analysis to single-$\vec{q}$ states, $\Delta_{\vec{q}} \propto \delta_{\vec{q},\vec{q}_0}$, the value of $\vec{q}_0$ is determined by the minimum of $a^{\text{S}}_{\vec{q}}$.  We define the critical current $J_c(\hat{\vec{n}})$ along $\hat{\vec{n}}$ as the maximal magnitude of $\vec{J}(\vec{q}) = 2e \Delta_{\vec{q}} \vec{\nabla} a^{\text{S}}_{\vec{q}}$ oriented along $\hat{\vec{n}}$. Postponing a microscopic evaluation of $a^{\text{S}}_{\vec{q}}$ to \secref{ModelStudies} below, we here only note that a unitary symmetry with vector representation $g$ implies $a^{\text{S}}_{\vec{q}} = a^{\text{S}}_{g\vec{q}}$ and, thus, $\vec{J}(g\vec{q}) = g\vec{J}(\vec{q})$ as well as $J_c(\hat{\vec{n}}) = J_c(g\hat{\vec{n}})$.

To quantify the `degree' or amount of diode effect, it will be convenient for us to first introduce a signed efficiency parameter 
\begin{equation}
    \eta_s(\hat{\vec{n}}) := \frac{J_c(\hat{\vec{n}})-J_c(-\hat{\vec{n}})}{J_c(\hat{\vec{n}})+J_c(-\hat{\vec{n}})} \label{AngularResSDEEfficiency}
\end{equation}
such that the angle-resolved diode-effect efficiency is $\eta(\hat{\vec{n}})=|\eta_s(\hat{\vec{n}})|\geq 0$. 
We next discuss symmetry constraints on $\eta(\hat{\vec{n}})$. From the discussion above, it follows that a unitary symmetry with vector representation $g$ enforces $\eta_s(g \hat{\vec{n}}) = \eta_s(\hat{\vec{n}})$. Since $\eta_s(\hat{\vec{n}}) = -\eta_s(-\hat{\vec{n}})$, which immediately follows from the definition (\ref{AngularResSDEEfficiency}), the SDE efficiency is required to vanish along any direction $\hat{\vec{n}}_0$ obeying $g\hat{\vec{n}}_0 = - \hat{\vec{n}}_0$; in short:
\begin{equation}
    g\hat{\vec{n}}_0 = - \hat{\vec{n}}_0\quad \Rightarrow \quad \eta(\hat{\vec{n}}_0)=0. \label{UnitarySymConstraint}
\end{equation}
Clearly, \equref{UnitarySymConstraint} is obeyed for all $\hat{\vec{n}}$ if $g$ is inversion ($I$) or two-fold rotation along the out-of-plane ($\hat{\vec{z}}$) direction $C_{2z}$ for a 2D system; these symmetries are thus required to be broken to allow for a finite SDE. 
While the following constraints are readily generalized to three spatial dimensions, we first focus for notational simplicity on a 2D system in the $xy$ plane. If $g$ is a two-fold rotational symmetry, $C_{2\hat{\vec{m}}}$, along an in-plane direction $\hat{\vec{m}}$ or a mirror-plane symmetry, $\sigma_{\hat{\vec{m}}}$, spanned by $\hat{\vec{z}}$ and $\hat{\vec{m}}$, \equref{UnitarySymConstraint} implies that $\eta(\hat{\vec{n}})=0$ for $\hat{\vec{n}}$ perpendicular to $\hat{\vec{m}}$. In other words, the zeros of $\eta(\hat{\vec{n}})$ are pinned along high-symmetry directions if the point group contains two-fold rotations along in-plane directions or mirror planes perpendicular to the system plane. Note that rotations with degrees higher than two do not pin the zeros.

Similar constraints arise from magnetic point symmetries, i.e., non-unitary symmetries, $\Theta g_m$, given by the product of time-reversal $\Theta$ and some unitary symmetry $g_m$. Such a magnetic symmetry implies $a^{\text{S}}_{\vec{q}} = a^{\text{S}}_{-g_m\vec{q}}$ and consequently $J_c(\hat{\vec{n}}) = J_c(-g_m\hat{\vec{n}})$; this, in turn, leads to $\eta_s(g_m \hat{\vec{n}}) = \eta_s(-\hat{\vec{n}}) = -\eta_s(\hat{\vec{n}})$. As such, the presence of the magnetic symmetry $\Theta g_m$ implies that
\begin{equation}
    g_m\hat{\vec{n}}_0 =  \hat{\vec{n}}_0\quad \Rightarrow \quad \eta(\hat{\vec{n}}_0)=0.   \label{MagneticSymConstraint}  
\end{equation}
Equation (\ref{MagneticSymConstraint}) contains the special case of preserved time-reversal symmetry, $g_m=\mathbbm{1}$ for which $g_m \hat{\vec{n}} = \hat{\vec{n}}$ always holds, implying $ \eta(\hat{\vec{n}}) = 0$ for all $\hat{\vec{n}}$; this just recovers the often-stated fact that $\Theta$ needs to be broken to get a SDE. However, \equref{MagneticSymConstraint} also shows that zeros of the angle-resolved SDE efficiency are pinned along high-symmetry directions if the system exhibits a magnetic reflection symmetry, $\Theta \sigma_{\hat{\vec{m}}}$, or magnetic rotational symmetry, $\Theta C_{2\hat{\vec{m}}}$; in both cases, it holds $\eta(\hat{\vec{m}})=0$. We will see examples of all of these constraints in the explicit model calculations of \secref{ModelStudies} below. 

We finally point out that \equsref{UnitarySymConstraint}{MagneticSymConstraint} also apply for the three-dimensional case. As such, we see that a two-fold non-magnetic rotation (mirror) symmetry lead to line (point) nodes for $\eta(\hat{\vec{n}})$ on the sphere $\{\hat{\vec{n}}\in\mathbb{R}^3|\hat{\vec{n}}^2=1\}$, while magnetic rotations (mirror) symmetries imply point (line) nodes.

\begin{figure}[tb]
   \centering
    \includegraphics[width=1.0\linewidth]{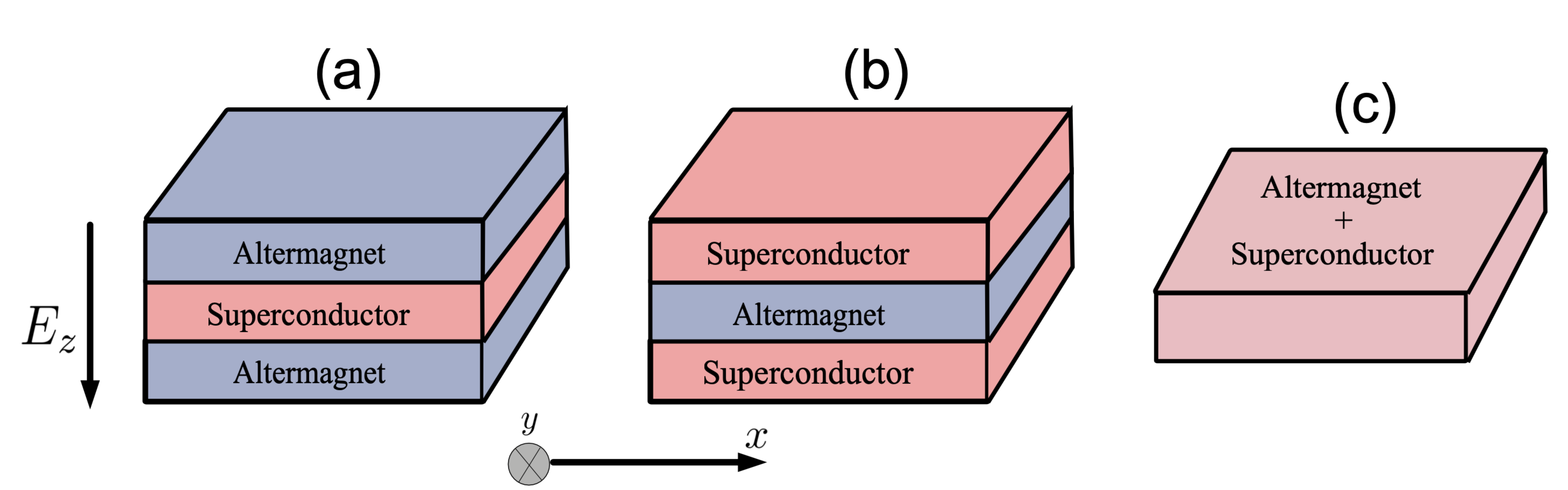}
    \caption{Schematic of the possible realizations of the considered models by (a) proximity-induced AM in a superconductor, (b) proximity-induced superconductivity on an AM metal, and (c) both orders coexisting in a single material. Our analysis also applies to the case where the proximitizing material, the AM in (a) and the superconductor in (b), is only present on one side. The advantage of the symmetric design is that an external electric field $E_{z}$ can be used to reduce the symmetries of the systems in a controllable way.}
    \label{fig:physicalrealisations}
\end{figure} 

\subsection{AM-induced SDE}
With these symmetry constraints at hand, we are now in position to analyze under which conditions an AM moment can induce a SDE and, if non-zero, whether the angle-resolved SDE efficiency $\eta(\hat{\vec{n}})$ has to have zeros along certain directions.  
To be precise, in this work, we use the following definition of an AM order parameter. We define an AM order parameter as a time-reversal-odd order parameter (i.e., some form of magnetic order), which unlike a ferromagnet does not lead to a finite net magnetic moment or spin-polarization; in addition, we require that it preserve lattice translation, unlike an antiferromagnet or more complex translational-symmetry-breaking orders (such as an incommensurate spiral or a tetrahedral magnet). 

For concreteness, we study the prototypical case of a minimal model for the normal state---one that only contains a single, spinful band,
\begin{equation}
    \mathcal{H}_0 = \sum_{\vec{k}} c^\dagger_{\vec{k},s} \left[ s_0 \epsilon_{\vec{k}} + \vec{g}_{\vec{k}} \cdot \vec{s} + \vec{N}_{\vec{k}} \cdot \vec{s} \right]_{s,s'} c^\pdagger_{\vec{k},s'},  \label{OneBandModel}
\end{equation}
where $c^\dagger_{\vec{k},s}$ are electron creation operators of spin $s$ and momentum $\vec{k}$, $\vec{s}=(s_x,s_y,s_z)$, $s_0 =  \mathbbm{1}$, are spin Pauli matrices. Furthermore, 
$\epsilon_{\vec{k}}$ is the bare dispersion, $\vec{g}_{\vec{k}}$ the spin-orbit vector, and $\vec{N}_{\vec{k}}$ the AM order parameter. Since these three quantities are, by design, even, even, and odd under $\Theta$, respectively, they have to obey
\begin{equation}
    \epsilon_{\vec{k}} = \epsilon_{-\vec{k}}, \qquad \vec{g}_{\vec{k}} = -\vec{g}_{-\vec{k}}, \qquad \vec{N}_{\vec{k}} = \vec{N}_{-\vec{k}}. \label{ConstraintsOnTheThreeTerms}
\end{equation}
We, therefore, conclude that $\epsilon_{\vec{k}}$ always preserves inversion, $\vec{g}_{\vec{k}}$ can only be non-zero if inversion, $I$, is broken (as usual), and the AM order parameter cannot break $I$ (and will always be odd under $\Theta I$) in the class of models in \equref{OneBandModel}. To describe superconductivity, we assume that Cooper pairs form in the spin-singlet channel, the superconducting order parameter transforms trivially under all point symmetries and preserves time-reversal.
Therefore, $\vec{N}_{\vec{k}}$ is the only term breaking time-reversal and $\vec{g}_{\vec{k}}$ the only source of broken inversion symmetry in \equref{OneBandModel}; this is why both are required to obtain a finite SDE, $\eta \neq 0$.

The different possible physical realizations of this class of models are illustrated in \figref{fig:physicalrealisations}. Either the AM term $\vec{N}_{\vec{k}}$ will be proximity-induced in superconducting systems, see \figref{fig:physicalrealisations}(a), or vice versa---a proximitized superconductor induces pairing in a metallic AM material [\figref{fig:physicalrealisations}(b)]. 
Although not essential for the realization of the following effects, we here assumed that the proximitizing materials are placed symmetrically on both sides such that an electric field, $E_z$, can be used to further reduce the symmetries \textit{in situ}. Our symmetry analysis and minimal-model study here still applies when the proximitizing material is only present on one side. In that case, the terms induced by $E_z \neq 0$ in the models below will already be present for $E_z=0$.  
Finally, our analysis also applies if both AM and superconductivity arise spontaneously in the same electron liquid, see \figref{fig:physicalrealisations}(c). We note in passing that this is the only scenario which would allow to realize altermagnetic diodes as a bulk effect in three dimensions and where the backaction mechanism of \refcite{banerjee_enhanced_2024} can apply, possibly giving rise to enhanced $\eta$.

To address the question which AM order parameters can induce a SDE, we start by considering all possible point groups $\mathcal{G}_0$ of 2D crystalline systems. For each of them, we go through all of its irreducible representations and check whether a generic time-reversal-odd $\vec{N}_{\vec{k}} \cdot \vec{s}$ in \equref{OneBandModel} transforming under it is an AM; this requires the presence of a residual symmetry that ensures that the expectation value of the total spin,
\begin{equation}
    \vec{S} = \sum_{\vec{k}} c^\dagger_{\vec{k},s} \vec{s}_{s,s'} c^\pdagger_{\vec{k},s'}, \label{SpinOperator}
\end{equation}
has to vanish. For instance, an AM order parameter transforming under $A_g$ of $\mathcal{G}_0 =  D_{2h}$ will lead to a residual $C_{2z}$ rotation symmetry, implying that $\braket{S_x}=\braket{S_y}=0$. The additional $\sigma_v$ mirror symmetry further implies that also $\braket{S_z}$ vanishes. In the first three columns of \tableref{PossibleAltermagnetsAndDiodes}, we list all such combinations of $\mathcal{G}_0$ and AM order parameters, some of which were already tabulated in \refcite{RafaelsPaper}. To organize the list, the first 11 lines refer to the AM order parameters for centrosymmetric point groups $\mathcal{G}_0$; interestingly, we find that, in all cases, these AM order parameters will continue to be proper AMs with $\braket{\vec{S}} = 0$ when an electric field is applied, $E_z \neq 0$, which reduces the point groups to $\mathcal{G}_{E_z} \leq \mathcal{G}_0$. Coming back to our previous example, $\mathcal{G}_0=D_{2h}$ will be reduced to $\mathcal{G}_{E_z} = C_{2v}$. It still contains the two symmetries used above to show that there is no finite magnetization.
The remaining lines of \tableref{PossibleAltermagnetsAndDiodes} refer to the AM order parameters in the 2D non-centrosymmetric point groups that are not reached by applying $E_z$ in centrosymmetric point groups. In all those cases, reducing the point group further by application of $E_z$ leads to a finite magnetization.

For all of these AMs and point groups, we investigate whether a superconducting order parameter in the singlet channel and transforming under the trivial representation of $\mathcal{G}_0$ can give rise to a SDE. Clearly, due to inversion symmetry, there cannot be a SDE for the centrosymmetric point groups at $E_z=0$. However, applying an electric field does allow for finite $\eta$ in a few cases, see penultimate column in \tableref{PossibleAltermagnetsAndDiodes}. This is remarkable since it shows that finite magnetization is not required to obtain a SDE. Also some of the AMs in certain non-centrosymmetric point groups allow for a SDE without finite moment. In \tableref{PossibleAltermagnetsAndDiodes} we also list whether and, if so, how the residual symmetries pin the zeros of $\eta(\hat{\vec{n}})$ to high-symmetry lines. These aspects will be further discussed on a case-by-case basis in the following section.

\begin{table*}[tb]
\begin{center}
\caption{Summary of possible AM order parameters $\vec{N}_{\vec{k}}$ in 2D models of the form of \equref{OneBandModel}. For each point group $\mathcal{G}_0$, we list all the irreducible representations (IR) that can give rise to an AM; the respective form of $\vec{N}_{\vec{k}}$ for small $\vec{k}$ is shown in the third column, where $\tau$ and $R_{\varphi}$ are, respectively, a real-valued constant and a rotation matrix along the third direction not fixed by symmetry. Applying an electric field $E_z$ reduces the point group to $\mathcal{G}_{E_z}$. In case of the centrosymmetric $\mathcal{G}_0$ (first 11 lines), all order parameters remain AMs (as indicated in the fourth column), characterized by zero net magnetization; this is not the case for the non-centrosymmetric $\mathcal{G}_0$ (line 12-20). In the column SDE, we indicate whether the respective AM will give rise to an SDE (in the centrosymmetric point groups, only possible for $E_z \neq 0$). We also indicate whether the angle-resolved SDE efficiency $\eta(\hat{\vec{n}})$ has zeros along generic directions $\hat{\vec{n}}_0$ (and their $C_{3z}$-related partners) or along crystalline axes, $\hat{\vec{e}}_{x,y}$. For $D_3$ and $D_{6}$, we introduced $\hat{\vec{n}}^i_{E_z}$ with  $\hat{\vec{n}}^i_{E_z=0} = \hat{\vec{e}}_i$, $i=x,y$, and pointing along a generic direction for $E_z \neq 0$.
The last column shows candidate materials, based on their conduction type---metallic (M) or insulating (I)---focusing on those cases where an SDE is present. }
\label{PossibleAltermagnetsAndDiodes}
\begin{ruledtabular}
 \begin{tabular}{cccccc} 
$\mathcal{G}_0$ & IR of AM & $\vec{N}_{\vec{k}}$ & $\mathcal{G}_{E_z}$ & SDE & Materials
\\ \hline
$C_{4h}$ & $B_{g}$ & $(0,0,k_{x}^2-k_{y}^{2}+\tau k_{x}k_{y})$ & $C_{4}$;  AM & \xmark \\ \hline
$C_{6h}$ &\begin{tabular}[c]{@{}c@{}}$B_{g}$\end{tabular} & \begin{tabular}[c]{@{}c@{}}$R_{\varphi}(k_{x}^2-k_{y}^{2}, -2k_{x}k_{y},0)$\end{tabular} & \begin{tabular}[c]{@{}c@{}} $C_{6}$;  AM \end{tabular} & \begin{tabular}[c]{@{}c@{}} $\eta(\pm C_{3z}^j\hat{\vec{n}}_0)=0$\end{tabular} \\ \hline
$D_{2h}$ & $A_{g}$ & $(0,0,k_x k_y)$ & $C_{2v}$;  AM & \xmark \\ \hline
$D_{4h}$ & \begin{tabular}[c]{@{}c@{}}$A_{1g}$\\$B_{1g}$\\$B_{2g}$\end{tabular} & \begin{tabular}[c]{@{}c@{}}$(0,0,k_{x}k_{y}(k_{x}^2-k_{y}^{2}))$\\$(0,0,k_{x}k_{y})$\\$(0,0,k_x^{2}-k_y^{2})$\end{tabular} & $C_{4v}$;  AM & \begin{tabular}[c]{@{}c@{}}\xmark\\\xmark\\\xmark\end{tabular} \\ \hline
$D_{6h}$ & \begin{tabular}[c]{@{}c@{}}$A_{1g}$\\$B_{1g}$\\$B_{2g}$\\$E_{2g}$\end{tabular} & \begin{tabular}[c]{@{}c@{}}$(0,0,k_{x}k_{y}(k_{x}^2-3k_{y}^{2})(3k_{x}^2-k_{y}^{2}))$\\$(k_{x}^2-k_{y}^{2},-2k_{x}k_{y},0)$\\$(2k_{x}k_{y},k_{x}^2-k_{y}^{2},0)$\\$(0,0,2k_{x}k_{y}), (0,0,k_{x}^2-k_{y}^2)$\end{tabular} & $C_{6v}$;  AM & \begin{tabular}[c]{@{}c@{}}\xmark\\ $\eta(\pm C_{3z}^{j}\hat{\vec{e}}_{x})=0$\\ $\eta(\pm C_{3z}^{j}\hat{\vec{e}}_{y})=0$\\\xmark\end{tabular}  &\begin{tabular}[c]{@{}c@{}}CrSb (M)\cite{smejkal_beyond_2022}, MnTe (I)\cite{mnte_realisation,Mnte_hall_effect} \\ \\ \end{tabular} \\ \hline
$D_{3d}$ & \begin{tabular}[c]{@{}c@{}}$A_{1g}$\end{tabular} & \begin{tabular}[c]{@{}c@{}}$ (k_{x}^2-k_{y}^{2},-2k_{x}k_{y},\tau k_xk_y(k_x^2-3k_y^2)(k_y^2-3k_x^2))$ \end{tabular} & $C_{3v}$;  AM & \begin{tabular}[c]{@{}c@{}}  $\eta(\pm C_{3z}^{j}\hat{\vec{e}}_{y})=0$ \end{tabular} & Fe$_{2}$O$_{3}$, CoF$_{3}$ (I) \cite{smejkal_beyond_2022}\\ \hline
$D_{2}$ & $A$ & $(0,0,k_{x}k_{y})$ & $C_2$; no AM & \xmark \\ \hline
$D_{3}$ & \begin{tabular}[c]{@{}c@{}}$A_{1}$\end{tabular} & \begin{tabular}[c]{@{}c@{}} $ (k_{x}^2-k_{y}^{2},-2k_{x}k_{y},\tau k_xk_y(k_x^2-3k_y^2)(k_y^2-3k_x^2))$ \end{tabular} & $C_{3}$; no AM &  $\eta(\pm C_{3z}^{j}\hat{\vec{n}}_{E_z}^y)=0$   \\ \hline
$D_{4}$ & \begin{tabular}[c]{@{}c@{}}$A_{1}$ \\$B_{1}$\\ $B_{2}$\end{tabular} & \begin{tabular}[c]{@{}c@{}}$(0,0,k_{x}k_{y}(k_{x}^{2}-k_{y}^{2}))$\\$(0,0,k_{x}k_{y})$\\ $(0,0,k_{x}^{2}-k_{y}^{2})$\end{tabular} & $C_{4}$; no AM& \xmark \\ \hline
$D_{6}$ & \begin{tabular}[c]{@{}c@{}}$A_{1}$\\$B_{1}$\\ $B_{2}$\\$E_{2}$\end{tabular} & \begin{tabular}[c]{@{}c@{}}$(0,0,k_{x}k_{y}(k_{x}^{2}-3k_{y}^{2})(3k_{x}^{2}-k_{y}^{2}))$ \\$(k_{x}^{2}-k_{y}^{2},-2k_{x}k_{y},0)$\\ $(2k_{x}k_{y},k_{x}^{2}-k_{y}^{2},0)$\\$(0,0,2k_{x}k_{y}), (0,0,k_{x}^2-k_{y}^2)$\end{tabular} & $C_{6}$; no AM & \begin{tabular}[c]{@{}c@{}}\xmark\\$\eta(\pm C_{3z}^{j}\hat{\vec{n}}^y_{E_z})=0$  \\ $\eta(\pm C_{3z}^{j}\hat{\vec{n}}_{E_z}^x)=0$ \\ \xmark \end{tabular}&\begin{tabular}[c]{@{}c@{}}VNb$_{3}$S$_{6}$ (M)\cite{smejkal_beyond_2022,vnb3s6} \\ \\ \end{tabular} \\ 
 \end{tabular}
 \end{ruledtabular}
\end{center}
\end{table*}

\section{Model studies}\label{ModelStudies}
In order to demonstrate the features we have deduced by symmetry analysis above more explicitly and to discuss additional aspects, we here construct and solve concrete lattice models; to this end, we use three different point groups and AM order parameters that can induce an SDE as examples.

\subsection{General formalism}
In all of these lattice models, superconductivity will be described in the same way. We start with an attractive ($g>0$) interaction in the spin-singlet Cooper channel,  
\begin{equation}
    \mathcal{H}_I = -g \sum_{\vec{q}} C^\dagger_{\vec{q}} C^\pdagger_{\vec{q}}, \quad C^\pdagger_{\vec{q}} = \sum_{\vec{k}} c_{\vec{k}+\vec{q}/2} i s_y c_{-\vec{k}+\vec{q}/2},
\end{equation}
and add it to the Hamiltonian in \equref{OneBandModel}, $\mathcal{H}_0 \rightarrow \mathcal{H}_0 + \mathcal{H}_I$. We then perform a mean-field decoupling in the Cooper channel and introduce $\Delta_{\vec{q}} = g \braket{C_{\vec{q}}}$, leading to the following Hamiltonian
\begin{align}\begin{split}
        \mathcal{H}_{\text{MF}} = &\sum_{\vec{k}} c_{\vec{k}}^\dagger h_{\vec{k}} c_{\vec{k}}^\pdagger + \sum_{\vec{k},\vec{q}} \left[ \Delta_{\vec{q}} c^\dagger_{\vec{k}+\vec{q}/2} i s_y c^\dagger_{-\vec{k}+\vec{q}/2} + \text{H.c.} \right] \\&+ \sum_{\vec{q}} \frac{|\Delta_{\vec{q}}|^2}{g}, \quad h_{\vec{k}} = s_0\epsilon_{\vec{k}} + \vec{g}_{\vec{k}} \cdot \vec{s} + \vec{N}_{\vec{k}} \cdot \vec{s}. \label{MFHamtilonian}
\end{split}
\end{align}
Using a redundant Nambu basis, defined by $\Psi_{\vec{k}} = ( c_{\vec{k}+\vec{q}/2},i s_y c^\dagger_{-\vec{k}+\vec{q}/2})^{T}$, we obtain a Bogoliubov-de-Gennes Hamiltonian,
    \begin{equation}
 H_{\text{BdG}} = \begin{pmatrix}
        h_{\vec{k}+\vec{q}/2} & \Delta_{\vec{q}} \\ \Delta^*_{\vec{q}} & - s_y h^*_{-\vec{k}+\vec{q}/2} s_y
    \end{pmatrix}  \label{BdGHamiltonian}
\end{equation}
which allows us to rewrite the Hamiltonian \eqref{MFHamtilonian} for a given $\vec{q}$ as,
\begin{equation}
    \mathcal{H}_{\text{MF}}(\vec{q}) = \sum_{\vec{k}} \Psi_{\vec{k}}^\dagger H_{\text{BdG}}\Psi_{\vec{k}}^\pdagger +  \frac{|\Delta_{\vec{q}}|^2}{g}.
    \label{eq:BdG}
\end{equation}
Next, we integrate out the electrons in this quadratic mean-field Hamiltonian and recover the Ginzburg-Landau expression in \equref{eq:gl_free}. The key quantity $a^{\text{S}}_{\vec{q}}$ is related to the particle-particle bubble $\Gamma(\vec{q})$ at finite $\vec{q}$ in the usual way, $a^{\text{S}}_{\vec{q}}= g^{-1}-\Gamma(\vec{q})$. Keeping both intra- and inter-band contributions, the particle-particle bubble is of the form
\begin{equation}
    \Gamma(\vec{q}) = \frac{1}{4}\sum_{\vec{k},\nu = p,m,\eta=\pm}   \Lambda_{\vec{k},\vec{q},\nu,\eta}\tanh{\frac{\mathcal{E}_{\vec{k},\vec{q},\nu,\eta}}{2T}},
\end{equation}
where $\Lambda_{\vec{k},\vec{q},\nu,\eta}$ contains the coherence factors and $\mathcal{E}_{\vec{k},\vec{q},\nu,\eta}$ are Bogoliubov energies (see \appref{latticeharmonics} for their explicit form); the indices $\nu$ and $\mu$ label the four different states associated with spin and particle-hole space in our redundant Nambu basis, cf.~\equref{BdGHamiltonian}. From the resulting $a^{\text{S}}_{\vec{q}}$, we compute the directional dependence of the critical current and hence $\eta(\hat{\vec{n}})$ as described in \secref{diode_efficiency}.
We will next apply this formalism using a few different normal-state Hamiltonians, $\mathcal{H}_0$, to elucidate the nature of the SDE in various point groups.

\subsection{Pinned zeros and $E_z$ tunability in $D_{6h}$}\label{CaseD6h}
We start from the centrosymmetric point group $\mathcal{G}_0=D_{6h}$, where $\epsilon_{\vec{k}}$ obeys $\epsilon_{\vec{k}} = \epsilon_{-\vec{k}} = \epsilon_{C_{3z}\vec{k}} = \epsilon_{\sigma_v\vec{k}}$. To model such a dispersion, we take a nearest-neighbor hopping model on the triangular lattice, $\epsilon_{\vec{k}} = - \sum_{j=1}^3 t \cos (\vec{a}_j\cdot \vec{k})$, where $\vec{a}_j$ are three $C_{3z}$-related primitive vectors. For notational simplicity, we measure all energies in units of $t$ in the following. The presence of $I$ in $D_{6h}$ implies $\vec{g}_{\vec{k}}=0$. We can see in \tableref{PossibleAltermagnetsAndDiodes} that $D_{6h}$ has four possible AM order parameters and only two of them---those transforming under $B_{1g}$ and $B_{2g}$---can give rise to a SDE when also applying an electric field $E_z$. For concreteness, we will here focus on $B_{1g}$, which reduces $D_{6h}$ to the magnetic point group generated by $C_{3z}$, $\Theta C_{2z}$, $\Theta\sigma_v$, and $I$. We construct the lowest-order, Brillouin-zone-periodic form of  $\vec{N}_{\vec{k}}$ consistent with $B_{1g}$, which corresponds to first-nearest-neighbor-hopping processes on the triangular lattice, with explicit form presented in \appref{latticeharmonics}; it behaves as
\begin{equation}
    \vec{N}_{\vec{k}} \sim N_{0} (k_x^2-k_y^2, -2 k_x k_y,0)^T \label{FirstExplicitFormofAMOP}
\end{equation}
around the $\Gamma$ point, i.e., for $|\vec{k}| \rightarrow 0$. Due to $\vec{N}_{\vec{k}} = \vec{N}_{-\vec{k}}$, see \equref{ConstraintsOnTheThreeTerms}, and $B_{1g}$ being odd under $C_{2z}$, it holds $(\vec{N}_{\vec{k}})_z = 0$ to arbitrary order in $\vec{k}$. 
Note that this order parameter obeys all of the properties in our definition of an AM: by design [see \equref{MFHamtilonian}] it is translationally invariant and breaks time-reversal symmetry (due to $\vec{N}_{\vec{k}} = \vec{N}_{-\vec{k}}$). The absence of finite spin polarization immediately follows by noting that the residual $C_{3z}$ and $C_{2z}\Theta$ symmetries (also present for $E_z \neq 0$) imply, respectively, that $\braket{S_{x,y}}$ and $\braket{S_z}$ have to vanish, which we have also checked explicitly. 

\begin{figure}[tb]
   \centering
    \includegraphics[width=0.9\linewidth]{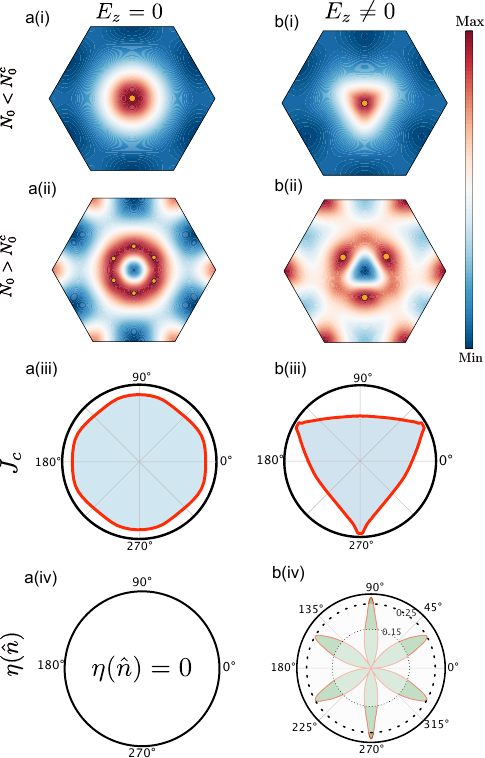}
    \caption{SDE in point group $D_{6h}$. Column (a) and (b) compare the cases for zero and non-zero electric field $E_z$. Row (i) and (ii) show the particle-particle bubble $\Gamma(\vec{q})$ in the Brillouin zone below and above the critical value of the strength, $N_0$, of the AM order; the orange circles denote the location of the maxima. Row (iii) shows the critical current  $J_c(\hat{\vec{n}})$ as a function of direction $\hat{\vec{n}}$. The resulting angle-resolved SDE efficiency (\ref{AngularResSDEEfficiency}) is plotted in the last row (iv) and is only finite for $E_z\neq 0$. The magnetic point group dictates the zeros of $\eta$ to be confined along high-symmetry directions, as discussed in the main text. Explicit parameters are given in \cite{Caption2}.}
    \label{fig:d6h}
\end{figure}

To understand the salient features of this model, let us start first with $E_z=0$, with results displayed in the first column (a) of \figref{fig:d6h}. One can see in \figref{fig:d6h}(ai) and (aii) that the maximum of $\Gamma(\vec{q})$ is pinned to $\vec{q}=0$ as long as $N_0$ in \equref{FirstExplicitFormofAMOP} is smaller than a critical value. This pinning is a consequence of $C_{3z}$ symmetry. It means that, in the current model, a sufficiently strong AM order is required to induce finite-momentum pairing. While finite-momentum pairing can be induced by AM order, as pointed out in recent publications in the context of different models without $C_{3z}$ symmetry \cite{zhang2023finitemomentum,chakraborty_zero-field_2023}, it does not necessarily follow immediately, as we can see here. Moreover, since we have inversion symmetry and unbroken $C_{3z}$, we observe that $\Gamma(\vec{q})$ is six-fold symmetric. This symmetry is inherited by the critical current $J_c(\hat{\vec{n}})$ shown in \figref{fig:d6h}(aiii); it thus obeys $J_c(\hat{\vec{n}}) = J_c(-\hat{\vec{n}})$ and there is no SDE, $\eta = 0$, see \figref{fig:d6h}(aiv).

As anticipated in \tableref{PossibleAltermagnetsAndDiodes}, we therefore have to apply an electric field $E_z$, which reduces $D_{6h}$ to $C_{6v}$, to get an SDE. On the level of the Hamiltonian, the electric field induces a non-zero spin-orbit vector $\vec{g}_{\vec{k}}$. We again refer to \appref{latticeharmonics} for its explicit form throughout the Brillouin zone and here only note that $\vec{g}_{\vec{k}} \sim \alpha (k_y,-k_x,0)^T$ as $|\vec{k}| \rightarrow 0$. Similar to $(\vec{N}_{\vec{k}})_z = 0$ noted above, the last component $(\vec{g}_{\vec{k}})_z$ has to vanish to all orders in $\vec{k}$ as a consequence of $C_{2z}$ and \equref{ConstraintsOnTheThreeTerms}. 

The corresponding results for finite $E_z$ are shown in the second column, (b), of \figref{fig:d6h}. We observe that, even in the presence of $\alpha \neq 0$, the maximum of $\Gamma(\vec{q})$ is pinned to zero momentum, see \figref{fig:d6h}(bi), due to the still unbroken $C_{3z}$ symmetry. However, as a consequence of inversion symmetry breaking, $\Gamma(\vec{q})$ is not an even function anymore. As soon as a critical value of $N_0$ is surpassed, the maxima of $\Gamma(\vec{q})$ move from zero $\vec{q}$ to three non-zero, $C_{3z}$-related momenta, as indicated by the orange circles in \figref{fig:d6h}(bii). The number of degenerate maxima is reduced from six to three [compared with \figref{fig:d6h}(aii)] when applying $E_z \neq 0$ as $I$ is broken. 

By the same token, the critical current $J_{c}(\hat{\vec{n}})$ is not an even function anymore, see \figref{fig:d6h}(biii), and we obtain a SDE, $\eta \neq 0$, as can be clearly seen in \figref{fig:d6h}(biv). This demonstrates explicitly that a SDE is also possible when the only source of broken $\Theta$ is an AM, i.e., in the absence of a net magnetic moment. As the SDE requires broken $I$, it is tunable by an electric field. 
What is more, we observe that $\eta(\hat{\vec{n}})$ has zeros along the high-symmetry directions $\hat{\vec{e}}_x$, $C_{3z}\hat{\vec{e}}_x$, and $C_{3z}^2\hat{\vec{e}}_x$ as follows from the magnetic reflection symmetry $\Theta \sigma_v$ and \equref{MagneticSymConstraint} [or $\sigma_d$ and \equref{UnitarySymConstraint} for that matter]. 
For an AM order parameter transforming under $B_{2g}$, the behavior is very similar. The main difference is that the aforementioned zeros of $\eta(\hat{\vec{n}})$ are pinned along the orthogonal directions $\hat{\vec{e}}_y$, $C_{3z}\hat{\vec{e}}_y$, and $C_{3z}^2\hat{\vec{e}}_y$, since, instead of $\Theta \sigma_v$, we have the magnetic reflection symmetry $\Theta \sigma_d$.

\subsection{Generic zeros in $C_{6h}$}
To demonstrate explicitly that the zeros of $\eta(\hat{\vec{n}})$ are not always pinned along high-symmetry directions for the SDE induced by an AM, we next consider the case $\mathcal{G}_0 = C_{6h}$. It exhibits no rotations $C_{2\hat{\vec{m}}}$ or reflections $\sigma_{\hat{\vec{m}}}$ with in-plane $\hat{\vec{m}}$, such that neither \equref{UnitarySymConstraint} nor \equref{MagneticSymConstraint} can apply (irrespective of the form of broken time-reversal symmetry). One consequence of the reduced symmetries is that the bare dispersion exhibits more terms compared to $D_{6h}$ discussed above, which we describe with the leading-order contribution, $\propto t'$, in 
\begin{equation}
    \epsilon_{\vec{k}} = - \sum_{j=1}^3 \left( t \cos{(\vec{a}_j\cdot \vec{k})} + t' \cos{(\vec{a}'_j\cdot \vec{k})}\right).
    \label{eq:c3z}
\end{equation}
Here, $\vec{a}'_j$ are three $C_{3z}$-related next-nearest-neighbor vectors on the triangular lattice ($\vec{a}'_{1} = 2\vec{a}_{2}-\vec{a}_{1}$, $\vec{a}'_{j} = C_{3z}^{j-1} \vec{a}'_{1}$). As can be seen in \tableref{PossibleAltermagnetsAndDiodes}, only one of the two possible AM order parameters---the one transforming under $B_g$ of $C_{6h}$---can give rise to a SDE. As a result of the reduced symmetry, the two different AM order parameters associated with $B_{1g}$ and $B_{2g}$ in $D_{6h}$ now both transform under $B_g$ of $C_{6h}$ and can, hence, mix. Introducing a $3\times 3$ matrix $R_{\varphi}$ describing rotations by $\varphi$ along the third direction, this admixture can be parametrized as
\begin{equation}
    \vec{N}_{\vec{k}} = R_{\varphi} \vec{N}^{B_{1g}}_{\vec{k}} \sim N_{0} R_{\varphi} (k_x^2-k_y^2, -2 k_x k_y,0)^T,
    \label{eq:AM_c6h}
\end{equation}
where $\vec{N}^{B_{1g}}_{\vec{k}}$ is the $B_{1g}$ AM order parameter of $D_{6h}$ used above in \secref{CaseD6h} and the asymptotic relation again refers to the vicinity of the $\Gamma$ point, $\vec{k} \rightarrow 0$.

\begin{figure}[tb]
   \centering
    \includegraphics[width=\linewidth]{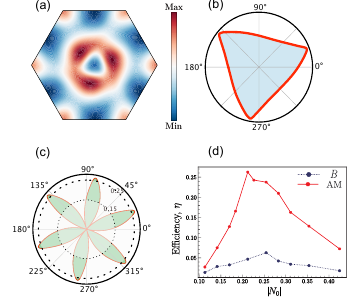}
    \caption{SDE in the point group $C_{6h}$ at $E_z \neq 0$. (a) Particle-particle bubble $\Gamma(\vec{q})$ in the Brillouin zone; (b) shows the critical current $J_c(\hat{\vec{n}})$ as a function of direction $\hat{\vec{n}}$ for $N_{0}=-0.21$; (c) displays the angle-resolved SDE efficiency $\eta(\hat{\vec{n}})$. Due to the reduced symmetry compared to $D_{6h}$ [cf.~\figref{fig:d6h}b(iv)], the zeros (and also maxima) of $\eta$ are rotated away from high-symmetry directions; (d) SDE efficiency $\eta$ as a function of magnetic field and altermagnetic order parameter $N_{0}$ shown in blue and red respectively. The magnetic field is chosen along the $x$ direction. In this figure, we used $\alpha=0.4, t'=0.05t$,  $g^{-1} = 0.5\Gamma(\vec{q}=0)$, $\varphi = -0.25 \pi$, $\varphi' = 0$.}
    \label{fig:c6h}
\end{figure}

To obtain a finite SDE, $\eta \neq 0$, inversion symmetry $I$ still needs to be broken by application of an electric field $E_{z} \neq 0$. This induces a spin-orbit coupling term $\vec{g}_{\vec{k}}$. For the same reason as above, the spin-orbit vector is less constrained than in $D_{6h}$ and can be written in the form $R_{\varphi'}\vec{g}_{\vec{k}}$, where $\vec{g}_{\vec{k}}$ is identical to the one used for $D_{6h}$ above, obeying $\vec{g}_{\vec{k}} \sim \alpha (k_y,-k_x,0)^T$. For $\vec{N}_{\vec{k}},\,\vec{g}_{\vec{k}} \neq 0$, the residual magnetic point group of the system is generated by $C_{3z}$ and $C_{2z}\Theta$. As such, the only constraints on the particle-particle bubble and the critical current are $\Gamma(\vec{q}) = \Gamma(C_{3z}\vec{q})$ and $J_c(\hat{\vec{n}}) = J_c(C_{3z}\hat{\vec{n}})$, which are clearly visible in our numerical results in \figref{fig:c6h}(a) and (b), respectively. As expected, the maxima and minima of these two quantities are not pinned along high-symmetry directions and, therefore, the zeros of $\eta(\hat{\vec{n}})$, shown in \figref{fig:c6h}(c), are not pinned either. 

We further present in \figref{fig:c6h}(d) the diode efficiency $\eta$ for both the magnetic field-induced SDE and the altermagnetic SDE. To treat them on equal footing, we use the same model and set $\vec{B}^{2} = N_{0}^{2} = \sum_{\vec{k}\in \text{BZ}}\vec{N}_{\vec{k}}^{2}$. Employing the same parameters as in the other parts of \figref{fig:c6h}, we find that the efficiency is significantly larger in the AM case; this demonstrates that the AM SDE is \textit{not} a subleading and, thus generally, weaker effect compared to the magnetic-field-induced one, contrary to what one might have expected since the average magnetization is required to vanish for an AM. However, we caution that both magnetic fields and AMs can yield sizeable SDE with relative magnitudes depending largely on the parameters chosen.

\subsection{$E_z$ tunable zeros and polarization in $D_3$}
We finally discuss the AM-induced SDE for the case where $\mathcal{G}_0$ is already a non-centrosymmetric point group. As such, an electric field will not be required to get a SDE. Nonetheless, as we will see, $E_z$ can still be used to tune the properties of the system in a non-trivial way. For concreteness, let us choose $\mathcal{G}_0 = D_3$. As can be seen in \tableref{PossibleAltermagnetsAndDiodes}, there is only a single AM order parameter, transforming under the trivial representation of the point group. While, to leading (quadratic) order in $\vec{k}$, it has the same form as $\vec{N}_{\vec{k}}$ in \equref{FirstExplicitFormofAMOP}, the subleading corrections are different. In particular, the reduced symmetries allow for a finite $(\vec{N}_{\vec{k}})_z$ component;
\begin{equation}
    \vec{N}_{\vec{k}} \sim N_{0}(k_{x}^2-k_{y}^{2},-2k_{x}k_{y}, \tau k_xk_y(k_x^2-3k_y^2)(k_y^2-3k_x^2)),
    \label{eq:AM_D3}
\end{equation}
where we only kept the leading term in every component and $\tau$ is a real parameter that is not determined by symmetry. The total spin polarization still vanishes, $\braket{\vec{S}} = 0$, as required for an AM, which follows from the preserved $C_{3z}$ ($\braket{S_{x,y}}=0$) and $C_{2x}$ ($\braket{S_z} = 0$) symmetries.

Interestingly, the spin-orbit vector has two different contributions, $\vec{g}_{\vec{k}} = \vec{g}_{\vec{k}}^{\text{I}} + \vec{g}_{\vec{k}}^{\text{II}}$ where $\vec{g}_{\vec{k}}^{\text{I}} \sim \beta (k_{x}, k_{y},k_{y}(3k_{x}^{2}-k_{y}^{2}))$ is already allowed by the point group $D_{3}$ without the need of an external electric field, and $\vec{g}_{\vec{k}}^{\text{II}} \sim \alpha (k_{y}, -k_{x},k_{x}(k_{x}^{2}-3k_{y}^{2}))$ is the spin-orbit coupling term induced by a perpendicular electric field $E_{z}$, i.e., $\alpha \propto E_z + \mathcal{O}(E_z^3)$. The full expressions for $\vec{N}_{\vec{k}}$, $\vec{g}_{\vec{k}}^{\text{I}}$, and $\vec{g}_{\vec{k}}^{\text{II}}$ we use are provided in \appref{latticeharmonics}. 
Finally, for the dispersion, we restrict ourselves just to a nearest-neighbor hopping term on the triangular lattice, $\epsilon_{\vec{k}} = - \sum_{j=1}^3 t \cos (\vec{a}_j\cdot \vec{k})$.

\begin{figure}[tb]
   \centering
    \includegraphics[width=\linewidth]{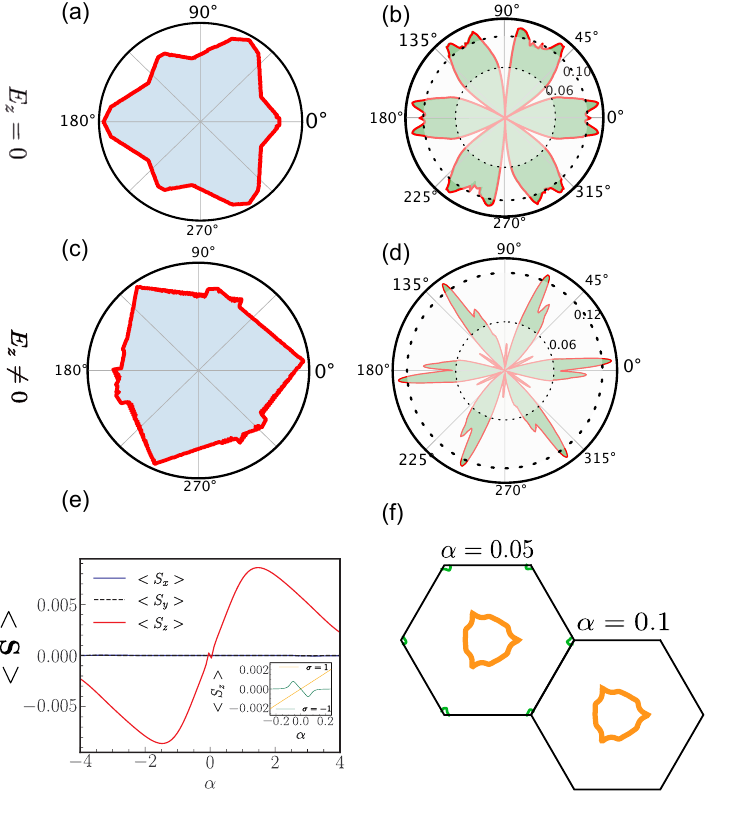}
    \caption{SDE in the non-centrosymmetric point group $D_{3}$. (a) and (c) show the directional-dependence of the critical current $J_c(\hat{\vec{n}})$ for zero and finite electric field, respectively. The corresponding SDE efficiencies $\eta(\hat{\vec{n}})$ are displayed in (b) and (c). Since $E_z \neq 0$ breaks all in-plane rotation symmetries $C_{2\hat{\vec{m}}}$, it can be used to rotate the symmetry-imposed zeros of $\eta(\hat{\vec{n}})$ away from high-symmetry directions. It can also be used, to control the net magnetization, see (e), where the spin polarization as a function of $E_z$ is shown; the inset is a zoomed-in version of the respective contribution of the two bands (in orange and green) to $\braket{S_z}$. (f) reveals that both Fermi surfaces (corresponding to the two bands) are present for $\alpha=0.05$ and, for $\alpha = 0.1$, only one of the bands contribute. Explicit parameters are given in \cite{d3}.}
    \label{fig:d3}
\end{figure}

Let us first focus on $E_z =0$ ($\alpha=0$). Since $C_{2z}$, $I$, and $\Theta$ are already broken, we find a finite SDE, see \figref{fig:d3}(a,b). Furthermore, we observe that the extrema of the critical current rotated by $90^{\circ}$ in comparison to $D_{6h}$ [cf.~\figref{fig:d6h}(biii)]. This is closely related to the discussion in \secref{diode_efficiency};  since $D_{3}$ has two-fold rotational symmetry $C_{2x}$ about the in-plane direction $x$, the current has to obey $J_{c}(C_{2x}\hat{\vec{n}}) = J_{c}(\hat{\vec{n}})$ which pins its extrema along $\hat{\vec{e}}_x$ and $C_{3z}$-related directions. Similarly, \equref{UnitarySymConstraint} implies that the angle-resolved efficiency $\eta(\hat{\vec{n}})$ has to go to zero for $\hat{\vec{n}}=\hat{\vec{e}}_{y}$, $C_{3z}\hat{\vec{e}}_{y} $ and $C_{3z}^{2}\hat{\vec{e}}_{y}$, which is also observed in our numerics [see \figref{fig:d3}(b)].

To demonstrate the effect of a finite $E_{z}$, we repeat the analysis now in the presence of the induced spin-orbit coupling term $\vec{g}_{\vec{k}}^{\text{II}}$. The electric field transforms as $A_{2}$ of $D_{3}$ and reduces the point group to $C_{3}$. Altogether, since time-reversal and inversion are still broken, the system still shows a SDE, see \figref{fig:d3}(c,d), only now that there is no two-fold in-plane rotation ($C_{2x}$) or a mirror-plane $\sigma_{\hat{\vec{m}}}$ as a symmetry. Naturally, it follows that the zeros of $\eta(\hat{\vec{n}})$ are not pinned to the high symmetry directions, but are now free to move in-plane, as is clearly visible in \figref{fig:d3}(d). An electric field can thus be used to tune the orientation of the zeros of the SDE efficiency $\eta(\hat{\vec{n}})$. 

By the same token, we also find that the $E_z$-induced reduction of the point group leads to the order parameter lose its AM nature---since all in-plane rotations are broken, we expect $\braket{S_z} \neq 0$ for $\alpha\neq 0$; meanwhile we still have $\braket{S_{x,y}}=0$ since $C_{3z}$ remains unbroken. This is indeed what we find when computing the expectation value of the spin operators in \equref{SpinOperator} explicitly within our model, with the result displayed in \figref{fig:d3}(e). We can further see that the orientation of the spin polarization can be swapped by changing the sign of $\alpha$, as expected by symmetry, and also by increasing $|\alpha|$ beyond a certain magnitude. 
To understand the latter observation better, we consider the inset in \figref{fig:d3}(e), which shows the contributions from the  bands $\sigma= \pm 1$ to the total polarization. As we start increasing $\alpha$ in the positive direction, for very small values of $\alpha$, the $\sigma=-1$ band  (shown in green) starts contributing more than its counterpart $\sigma=1$ (shown in orange) to $\braket{S_{z}}$. This happens until a certain value of $\alpha$ is reached from whence the $\sigma=1$ starts taking over and ends up contributing the bulk of the total polarization, thereby flipping the value of $\braket{S_{z}}$. This flipping of polarization on the same side can also be well understood by noting that at $\alpha = 0.05 t$, both Fermi surfaces [see \figref{fig:d3}(f)] corresponding to $\sigma=\pm 1$, shown in orange and green respectively contribute. Meanwhile, when $\alpha = 0.1 t$, only the green one contributes to the spin polarization. Taken together, we see that an AM order parameter in the non-certrosymmetric point group $D_3$ (same applies to $D_6$, see \tableref{PossibleAltermagnetsAndDiodes}) can induce a SDE, with both the zeros in the efficiency $\eta(\hat{\vec{n}})$ and the spin polarization  being tunable by an applied electric field $E_z$.

\section{Conclusion}\label{Conclusion}
In this work, we have studied under which conditions a conventional two-dimensional superconducting phase can exhibit a SDE, $\eta_s \neq 0$ in \equref{AngularResSDEEfficiency}, if the only source of broken time-reversal symmetry is an AM order parameter. To approach this question systematically, we started by listing all AM order parameters $\vec{N}_{\vec{k}}$, coupling to the electrons as given by \equref{OneBandModel}, for all possible 2D crystalline point groups, see \tableref{PossibleAltermagnetsAndDiodes}. For each of them, we checked whether there are still residual symmetries, such as inversion symmetry or two-fold rotation along a direction perpendicular to the plane of the system, that would enforce the critical currents along opposite directions to be the same, $J_c(\hat{\vec{n}}) = J_c(-\hat{\vec{n}})$. We found that there are indeed several point groups and associated AM order parameters ($C_{6h}$ with AM in $B_g$; $D_{6h}$ with AM in $B_{1g}$ and $B_{2g}$; $D_{3d}$ with $A_{1g}$; $D_{3}$ with AM in $A_1$; $D_6$ with $B_1$ and $B_2$) that give rise to an SDE, as we have also checked by explicit model calculations. Since, by design, the total magnetic moment vanishes when an AM is the only source of broken time-reversal symmetry, this demonstrates that the SDE does not require finite net magnetization. A direct comparison with the magnetic-field-induced SDE reveals that the altermagntic SDE is not a subleading effect; in fact, depending on parameters, the altermagnetic efficiency can even be substantially larger when compared in the same model with the magnetic-field scenario.

In the context of the SDE, the AM order parameters can be further categorized in the following way: the AMs transforming under $B_g$ of $C_{6h}$, $B_{1g}$ or $B_{2g}$ of $D_{6h}$, and $A_{1g}$ of $D_{3d}$ will only give rise to an SDE if also an electric field, $E_{z}$, perpendicular to the plane of the system is applied. This breaks inversion symmetry, a necessary requirement for the SDE. Note, however, that in all cases listed here, there are still enough symmetries present to guarantee a vanishing net magnetic moment. This shows that AMs allow for electric-field-tunable (rather than the frequently discussed magnetic-field-tunable) SDEs.
In turn, the AMs transforming under $A_1$ and $B_{1,2}$ for the non-centrosymmetric point groups $D_3$ and $D_6$, respectively, will immediately give rise to a SDE, without application of an electric field; moreover, for $E_z \neq 0$, the symmetries will be lowered in such a way that the order parameters cease to be AMs. Here the electric field can be used to control the out-of-plane magnetization of the system.

Another important distinction of the different scenarios for AM-induced SDEs can be made based on the directions $\hat{\vec{n}}_j$ where the SDE efficiency vanishes, $\eta (\hat{\vec{n}}_j) = 0$. While $\hat{\vec{n}}_j$ are pinned along high-symmetry directions for $D_{6h}$ (reduced to $C_{6v}$ by $E_z \neq 0$) and $D_{3d}$ (reduced to $C_{3v}$), this is not the case for the SDEs induced by the $B_g$ AM in $C_{6h}$ (reduced to $C_6$), where $\hat{\vec{n}}_j$ point along generic ($C_{3z}$-related) directions. Finally, for the AMs of non-centrosymmetric point groups $D_3$ and $D_6$ that can induce an SDE, applying an electric field $E_z \neq 0$ allows to rotate the $\hat{\vec{n}}_j$ with $\eta (\hat{\vec{n}}_j) = 0$ away from crystalline to generic directions.

These findings, that can be deduced from symmetry considerations, are summarized in \tableref{PossibleAltermagnetsAndDiodes}. We have further constructed and solved minimal lattice models for the different classes of and phenomena associated with AM-induced SDEs discussed above. This further illustrates and explicitly demonstrates the complex interplay of AM and superconductivity.
We finally point out that there are several setups that can be used to realize and probe the proposed AM-induced SDE experimentally, see \figref{fig:physicalrealisations}. Given that there are already candidate materials with the right symmetries (see last column in \tableref{PossibleAltermagnetsAndDiodes}), the rapidly growing number of AM candidate materials \cite{gao_ai-accelerated_2023}, and the fact that heterostructures are routinely grown or stacked, this seems to be well within current experimental reach.

\begin{acknowledgments}
S.B.~and M.S.S.~acknowledge funding by the European Union (ERC-2021-STG, Project 101040651---SuperCorr). Views and opinions expressed are however those of the authors only and do not necessarily reflect those of the European Union or the European Research Council Executive Agency. Neither the European Union nor the granting authority can be held responsible for them. The data sets generated in this study are available in the Zenodo database under the accession code \url{ https://zenodo.org/records/12794241}. The authors thank P.~McClarty, R.~Fernandes, L.~Pupim, A.~Rastogi and J.~Sobral for discussions.  
\end{acknowledgments}

\bibliography{draft_Refs}

\begin{thebibliography}{120}
\bibitem [{Cap()}]{Caption2}%
  \BibitemOpen
  \href@noop {} {\ }\bibinfo {note} {For a(i) weak altermagnetism
  $N_{0}=-0.1,\alpha=0$; a(ii) strong altermagnetism $N_{0}=-0.2,\alpha=0$;
  (b)(ii-iv) in presence of electric field $E_{z}$, $N_{0}=-0.23,\alpha=0.4$
  and $g^{-1} = 0.5\Gamma(\vec{q}=0)$. Note that $\alpha$ is proportional to
  both the intrinsic spin-orbit coupling and $E_z$.}\BibitemShut {Stop}%
\bibitem [{d3()}]{d3}%
  \BibitemOpen
  \href@noop {} {\ }\bibinfo {note} {For $E_z = 0$ we take $N_{0} = -0.1, \beta
  = 0.7, \alpha = 0$ and $g^{-1} = 0.5\Gamma(\vec{q}=0)$; for $E_{z}\neq 0$ in
  second row, $N_{0} = -1.4, \tau = 1, \beta = 1.2, \alpha = 1.0$ and $g^{-1} =
  0.8\Gamma(\vec{q}=0)$; (e), (f) $N_{0} = -1.4, \beta = 0.8$}\BibitemShut
  {NoStop}%
\end{thebibliography}

\onecolumngrid

\begin{appendix}

\section{Expression for the particle-particle bubble} \label{particleparticlebubble}
In this section, we state the expression for the particle-particle bubble in presence of both inter- and intraband pairings. 
Given the Hamiltonian, defined in Eq.\eqref{eq:BdG}, we begin by defining a few quantities,  
\begin{equation}
  l_{i,\vec{k},\vec{q}} = g_{i,\vec{k},\vec{q}} + N_{i,\vec{k},\vec{q}}, \quad i = x,y,z \quad   g_{p,\vec{k},\vec{q}}^{\pm} = l_{x,\vec{k}+\vec{q}/2} \pm i l_{y,\vec{k}+\vec{q}/2}, \quad      g_{m,\vec{k},\vec{q}}^{\pm} = l_{x,-\vec{k}+\vec{q}/2} \pm i l_{y,-\vec{k}+\vec{q}/2}.
\end{equation}
In addition the renormalized energies are given as
\begin{equation}
    \mathcal{E}_{\vec{k},\vec{q},p,\pm} = \epsilon_{\vec{k}+\vec{q}/2} \pm g_{p}, \quad
    \mathcal{E}_{\vec{k},\vec{q},m,\pm} = \epsilon_{-\vec{k}+\vec{q}/2} \pm g_{m}, \quad
\end{equation}
where $g_{p}^{2} = \sum\limits_{i=x,y,z} l_{i,\vec{k}+\vec{q}/2}^{2} $ and $g_{m}^{2} = \sum\limits_{i=x,y,z} l_{i,-\vec{k}+\vec{q}/2}^{2} $. 
Armed with these definitions, we can write down the various coherence factors $\Lambda_{\vec{k},\vec{q},\nu,\eta}$ for the spinful model as
\begin{equation}
    \Lambda_{\vec{k},\vec{q},m,\mp} = \mp\frac{G_{\vec{k},\vec{q}}}{g_{m}(\mathcal{E}_{\vec{k},\vec{q},m,\mp}+ \mathcal{E}_{\vec{k},\vec{q},p,+})(\mathcal{E}_{\vec{k},\vec{q},m,\mp}+ \mathcal{E}_{\vec{k},\vec{q},p,-})} + \frac{2}{(\mathcal{E}_{\vec{k},\vec{q},m,\mp}+ \mathcal{E}_{\vec{k},\vec{q},p,\mp})};
\end{equation}
\begin{equation}
    \Lambda_{\vec{k},\vec{q},p,\mp} = \mp\frac{G_{\vec{k},\vec{q}}}{g_{p}(\mathcal{E}_{\vec{k},\vec{q},m,-}+ \mathcal{E}_{\vec{k},\vec{q},p,\mp})(\mathcal{E}_{\vec{k},\vec{q},m,+}+ \mathcal{E}_{\vec{k},\vec{q},p,\mp})} + \frac{2}{(\mathcal{E}_{\vec{k},\vec{q},m,\mp}+ \mathcal{E}_{\vec{k},\vec{q},p,\mp})};
\end{equation}
where $G_{\vec{k},\vec{q}} = g_{m,\vec{k},\vec{q}}^{+} g_{p,\vec{k},\vec{q}}^{-} + g_{m,\vec{k},\vec{q}}^{-} g_{p,\vec{k},\vec{q}}^{+} + 2g_{p}g_{m}+ 2 l_{z,\vec{k}+\vec{q}/2}l_{z,-\vec{k}+\vec{q}/2}$.  
The final expression for the particle-particle bubble $\Gamma(\vec{q})$ then reads 
\begin{equation}
    \Gamma(\vec{q}) = \frac{1}{4}\sum_{\vec{k}}\left(\Lambda_{\vec{k},\vec{q},p,+}\tanh{\frac{\mathcal{E}_{\vec{k},\vec{q},p,+}}{2T}} +   \Lambda_{\vec{k},\vec{q},p,-}\tanh{\frac{\mathcal{E}_{\vec{k},\vec{q},p,-}}{2T}} + \Lambda_{\vec{k},\vec{q},m,+}\tanh{\frac{\mathcal{E}_{\vec{k},\vec{q},m,+}}{2T}} + \Lambda_{\vec{k},\vec{q},m,-}\tanh{\frac{\mathcal{E}_{\vec{k},\vec{q},m,-}}{2T}}\right).
\end{equation}
\section{Details of the models} \label{latticeharmonics}
 We here state the basis functions used to characterise the terms $\vec{N}_{\vec{k}}$ and $\vec{g}_{\vec{k}}$ of the Hamiltonian in Eq. \eqref{MFHamtilonian} of the main text for various point groups. Applying the convention, where one of the primitive lattice vectors of the underlying triangular lattice is given by $\vec{a}_1 = (1,0)^T$, the leading basis functions explicitly read as
\begin{figure}[H]
   \centering
    \includegraphics[width=0.8\linewidth]{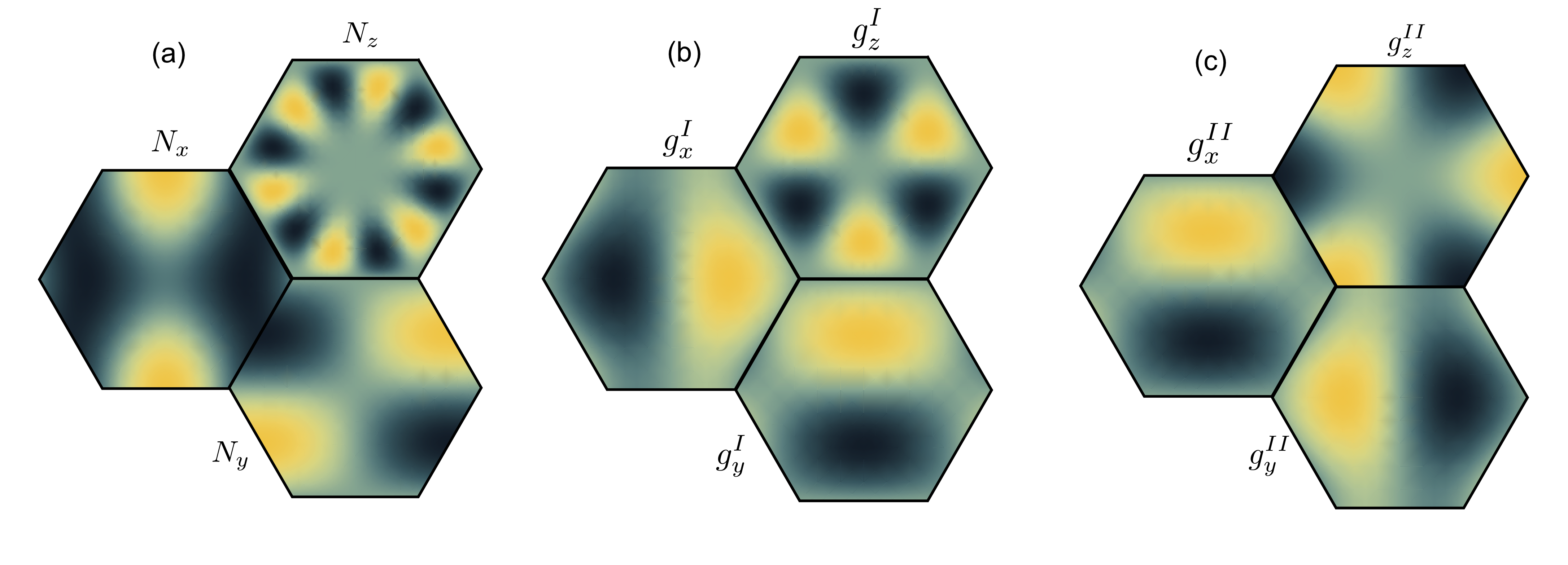}
    \caption{Multi-component momentum dependence of (a) the altermagnet term $\vec{N}_{\vec{k}}$; (b) inversion-symmetry breaking term $\vec{g}_{\vec{k}}^{I}$ and (c) shows the spin-orbit coupling term $\vec{g}_{\vec{k}}^{II}$ induced by $E_{z}$. For point groups $D_{6h}$ and $C_{6h}$, $\tau=0$, $\tilde{\tau}=0$ and $\beta=0$, while for $D_{3}$, $\tau\neq 0$, $\tilde{\tau}\neq 0$ and $\beta\neq0$. }
    \label{fig:lattice basis functions}
\end{figure}
\begin{equation}
    \vec{N}_{\vec{k}} = N_{0}\left(\frac{-8}{3}(\cos{k_x} - \cos{\frac{k_x}{2}} \cos{\frac{\sqrt{3}k_y}{2}}),\frac{-8\sin{\frac{k_x}{2}}\sin{\frac{\sqrt{3}k_y}{2}}}{\sqrt{3}},64\tau\frac{(\sin{\frac{5k_{x}}{2}}\sin{\frac{3\sqrt{3}k_{y}}{2}}+\sin{\frac{k_{x}}{2}}\sin{\frac{\sqrt{3}k_{y}}{2}}-\sin{2k_{x}}\sin{\sqrt{3}k_{y}})} {3 \sqrt{3}}\right)
    \label{eq:altermagnetbasis}
\end{equation}
\begin{equation}
\vec{g}^{\text{I}}_{\vec{k}}= \beta\left(\frac{2}{3}(2\cos{\frac{k_{x}}{2}} + \cos{\frac{\sqrt{3}k_{y}}{2}})\sin{\frac{k_{x}}{2}} \right., \left.{}\frac{2}{\sqrt{3}}\cos{\frac{k_{x}}{2}}\sin{\frac{\sqrt{3}k_{y}}{2}}, \frac{8(-2\cos{\frac{3k_{x}}{2}}\sin{\frac{\sqrt{3}k_{y}}{2}}+\sin{\sqrt{3}k_{y}})}{3\sqrt{3}}\right),
\label{eq:inversionbreakingbasis}
\end{equation}
\begin{equation}
\vec{g}^{\text{II}}_{\vec{k}}= \alpha\left(\frac{2}{\sqrt{3}}\cos{\frac{k_{x}}{2}}\sin{\frac{\sqrt{3}k_{y}}{2}} \right., \left.{}\frac{-2}{3}(2\cos{\frac{k_{x}}{2}} + \cos{\frac{\sqrt{3}k_{y}}{2}})\sin{\frac{k_{x}}{2}},  16\tilde{\tau} \left(-\cos{\frac{k_{x}}{2}} + \cos{\frac{\sqrt{3}k_{y}}{2}}\right)\sin{\frac{k_{x}}{2}} \right).
\label{eq:inducedspinorbitbasis}
\end{equation}
These are the functions with the least number of nodes/nodal lines that satisfy the required symmetry constraints while being periodic on the Brillouin zone. As explained in the main text,  $\tau = 0$ for point groups $D_{6h}$ and $C_{6h}$ and therefore Eq.~\eqref{eq:altermagnetbasis} reduces to Eq.~\eqref{FirstExplicitFormofAMOP} and Eq.~\eqref{eq:AM_c6h}, respectively, when expanded for small $\vec{k}$. As the point group $D_{3}$ permits the AM order parameter to also have a third component, it is given by Eq.~\eqref{eq:altermagnetbasis} with $\tau \neq 0$ resulting in Eq.~\eqref{eq:AM_D3} for $\vec{k}\rightarrow 0$. Similarly, the leading basis functions for the electric-field induced spin-orbit coupling $\vec{g}^{\text{II}}_{\vec{k}}$ are given by \equref{eq:inducedspinorbitbasis}, with $\tilde{\tau} =0$ for $\mathcal{G} = D_{6h}$ and $C_{6h}$; meanwhile $\tilde{\tau}\neq 0$ for $\mathcal{G} =D_{3}$. Finally, the leading spin-orbit vector at zero electric field for the point group $D_3$ is given by $\vec{g}^{\text{I}}_{\vec{k}}$ in \equref{eq:inversionbreakingbasis}.

\end{appendix}
\end{document}